\documentclass{caosp}

\usepackage{graphicx}

\articleNo{263}
\pubyear{2016}
\volume{46}
\volnumber{1}
\firstpage{5}
\received{July 12, 2016}
\accepted{Sept 5, 2016 }

\begin{document}

\htitle{Recent changes in a flickering variability of the black hole X-ray transient V616~Mon}
\hauthor{S.\,Shugarov, N.\,Katysheva, D.\,Chochol, N.\,Gladilina, E.\,Kalinicheva and  A.\,Dodin}

\title{Recent changes in a flickering variability of the black hole X-ray transient V616~Mon=A0620--00}

\author{
       S.\,Shugarov\inst{1,2}
      \and
       N.\,Katysheva\inst{2}
      \and
       D.\,Chochol\inst{1}
      \and
       N.\,Gladilina\inst{3}
       \and
       E.\,Kalinicheva\inst{4}
       \and
       A.\,Dodin\inst{2}
      }

\institute{
         \lomnica,
         \and
         Sternberg Astronomical Institute, Moscow State University,
         Universitetskij pr., 13, Moscow, 119991, Russia,
         \and
         Institute of Astronomy, Russian Academy of Sciences,
         Pyatnitskaya Str., 48, Moscow, 119017, Russia
         \and
         Faculty of Physics, Lomonosov Moscow State University, Leninskie Gory, Moscow, 119991, Russia
         }

\date{June 12, 2016}
\maketitle

\begin{abstract}
V616 Mon = A0620-00  is a prototype of black hole transient X-ray binaries. Our 2003-16 optical photometry of the object during X-ray quiescence, obtained by 50-250 cm telescopes in Crimea, Caucasus Mountains and Slovakia, consists of $\sim$ 7660 CCD frames in Johnson-Cousins $V,R,R_C,I$ bands and the integral light. During 2003, 2008-9 and 2015-16 passive states, the phase light curve of the binary exhibited mainly variations caused by an ellipsoidal shape of the red dwarf component. During 2004-6 and 2009-14 active states a significant aperiodic broad-band variability (flickering) was present, arising in a black hole accretion disk and a bright spot, where the mass transfer stream hits the outer edge of the disk. Long term photometry of our minima times, together with available positions of superior conjunctions of the red dwarf found from spectroscopy, allowed us to refine the orbital period of V616 Mon to 0.32301407(5) days.
\keywords{dwarf novae -- photometry, black hole, rapid variability}
\end{abstract}

%
\label{intr}

\section{Introduction}

A0620-00 was discovered on August 3, 1975 by the Ariel-5 satellite as a bright X-ray nova (Elvis et al., 1975) and designated as V616 Mon in the "General Catalogue of Variable Stars" (Kholopov et al., 1998). The precursor was detected on archival photographic plates at a magnitude of $B_{pg}\sim$  20.0-20.5 (Ward et al., 1975, Eachus et al., 1976).

The outburst light curves of the nova in X-ray, optical, infrared and radio wavelengths in 1975-76 were gathered by Kuulkers (1998). The nova reached 11.$^m$4 in the {\it B -} band (Shugarov, 1976), 30 Crab in X-rays and 200 mJy in the 962 MHz radio band. An outburst in X-rays was observed with satellites Ariel-5, SAS-3 and with the X-ray detector "Filin" on  the USSR orbital station SALYUT-4. The object was classified as a soft X-ray transient, and according to White et al. (1984), it was in a "high state" during most of the X-ray observations. The spectral changes during the rise to the outburst indicated a change from the "low state" to a high state. The different X-ray states (quiescent, low, intermediate, high, very high) depend on an increasing mass accretion rate through the accretion disk. After dramatic fading in the optical brightness in April 1976 (Martynov et al., 1976), V616 Mon passed into a quiescence (Lyuty \& Shugarov, 1979).

Eachus et al. (1976) found the previous outburst of the nova on archival plates of the Harvard College Observatory. The nova reached $B_{pg}=12^m$ in November 1917. Shugarov et al. (1976) detected the nova at two Simeiz plates taken on January 20, 1918 (JD 2421461.2) and estimated its brightness as $B_{pg}=12.^m4$. Two pairs of very short (0.5 - 5 ms) flares were detected in quiescence on February 13 and 14, 1986 by Shvartsman et al. (1989) with the 6-m BTA telescope. The brightness temperature in the flare was at least $10^9 - 10^{11}$K, indicating non-thermal accretion onto a black hole.

The blue-visual (4000-6400 \AA~FWHM) passband and $I$ photometry and spectroscopy of the object taken by McClintock \& Remillard (1986) showed that the object is a compact binary with the period of 0.3230141(4) days. The photometric binary cycle has a double hump structure as a result of the ellipsoidal shape of the red dwarf companion filling its Roche lobe. The authors derived a mass function $f(M)$ = 3.18$\pm$0.16 M$_\odot$ from the radial velocity curve of the red K5-K7 dwarf companion (Oke, 1977) and estimated the mass of the compact object to be $\geq$ 3.2 M$_\odot$, so it is an excellent black-hole candidate. McClintock \& Remillard (1986) determined the epoch of minimum light in the binary light curve (LC) at JD 244 5477.827. Using radial velocity measurements they established that this is the phase of an inferior conjunction of the compact object (the black hole is in front of the red dwarf with respect to the observer). The $R,I$ photometry in the years 1991-95 confirmed the presence of the double humped light curve with a period of  $P_{orb}=0^d.3230160$ (Leibowitz et al., 1998). While there were no indications of a short-term variability, the mean brightness of the object varied on a time-scale of hundreds of days. Subsequent $J,H,K$  photometry in 1999-2000, obtained by Gelino et al. (2001), has shown that the shape of an ellipsoidal light curve changed with time. The $V$ and infrared $I,H$ photometry spanning the period 1999-2007 was published by Cantrell et al. (2008). They distinguished three distinct optical states. Light curves in a "passive" state can be completely explained by periodic orbital variations and photometric errors. The "loop" and "active" states are characterized by the increases of scatter of observations. While a short time scale rms variability in the loop state was 0.05 mag in $V$ and 0.047 mag in $H$, in the active state it reached 0.088 mag in $V$ and 0.057 mag in $H$. The loop and active states are brighter than a passive state and were detected mainly in the years 2003-2007.

\begin{table}[t]
\small
\begin{center}
\caption{Journal of observations.}
\label{ts}
\begin{tabular}{llrclll}
\hline\hline
Yrs &{JD beginning-end}&  {N} &  {\footnotesize 1000}{$\cdot\sigma$} &  {System} & {Telescopes} & {Stage}\\
\hline
2003-04	  & 52965-53023 & 297 & 19 &  $V, R, I$ &125& passive\\
2004-05	  & 53317-53323 & 218 & 50 &  $V, R, I$ &125& active\\
2005-06	  & 53696-53710 &  45 & 54 &  $V, R$    &125& active\\
2008-09	  & 54891-54912 &  47 & 33 &  $R, R_C$   &60a, 50& passive\\
2009-10	  & 55095-55159 & 421 & 42 &  $R$     &60a   & active\\
2010-11	  & 55509-55517 &1389 & 69 &  $R$    &125    & active\\
2011-12	  & 55889-55910 & 362 & 39 &  $R$   &125     & active\\
2012-13	  & 56240-56252 &1048 & 92 &  $V$     &125    & active\\
2013-14	  & 56604-56611 &1450 & 74 &  $C$    &125, 60a & active\\
2014-15	  & 56958-56995 &1363 & 63 &  $C$    &125, 60b & active\\
2015-16	  & 57331-57479 &1021 & 18 &  $C, R_C, V$ &125, 250, 60b & passive\\
\hline\hline
\end{tabular}
\end{center}
~
{\footnotesize Telescopes: 50/200-cm in Star\'a Lesn\'a; 60/750-cm in Crimea (60a);
60/750-cm in Star\'a Lesn\'a (60b); 125/2000-cm in Crimea; 250/2000-cm in Caucasus.}
\end{table}

\begin{figure}
\centerline{\includegraphics[width=12cm,clip=]{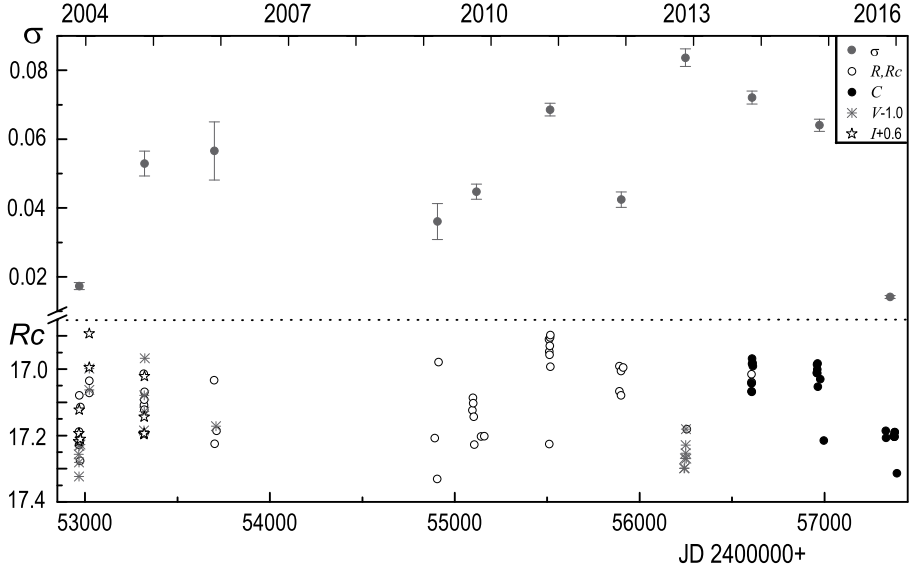}}
\vspace{-1mm}
\caption{Our 2003-16 light curve of V616 Mon in $V,R,R_C,I$ passbands and the integral light $C$. Every symbol is the mean night value of the magnitude in a particular passband, normalized to the $R_C$ magnitude scale. The $R,R_C$ and $C$ magnitudes are close to each other. The activity of the system is demonstrated by a standard deviation $\sigma$ of the observations taken during every observational season.}
\vspace{-3mm}
\end{figure}

\begin{figure}
\centerline{\includegraphics[width=11cm,clip=]{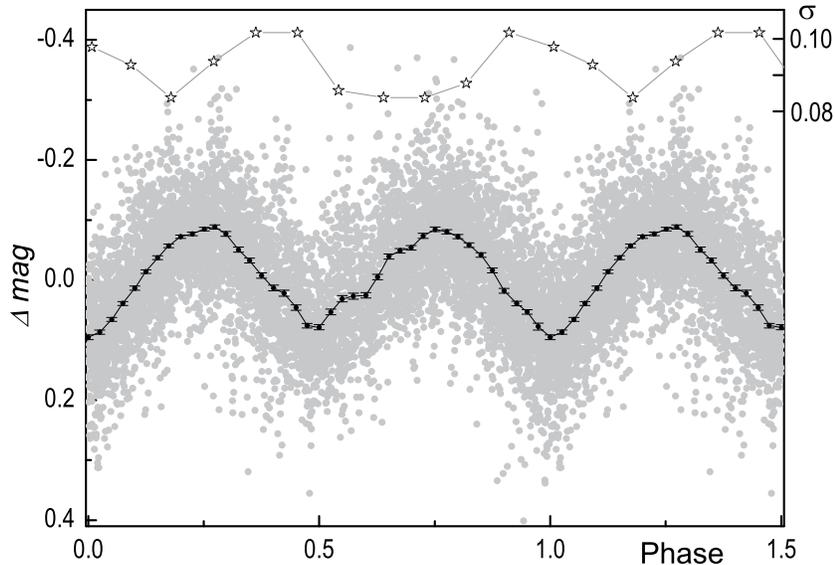}}
\caption{The LC of V616 Mon folded with the ephemeris (1) and dependence of a standard deviation $\sigma$ on the orbital phase (top).}
\end{figure}

\begin{figure}[h!]
\vspace{-2mm}
\begin{minipage}[h]{0.49\linewidth}
\includegraphics[width=1\linewidth]{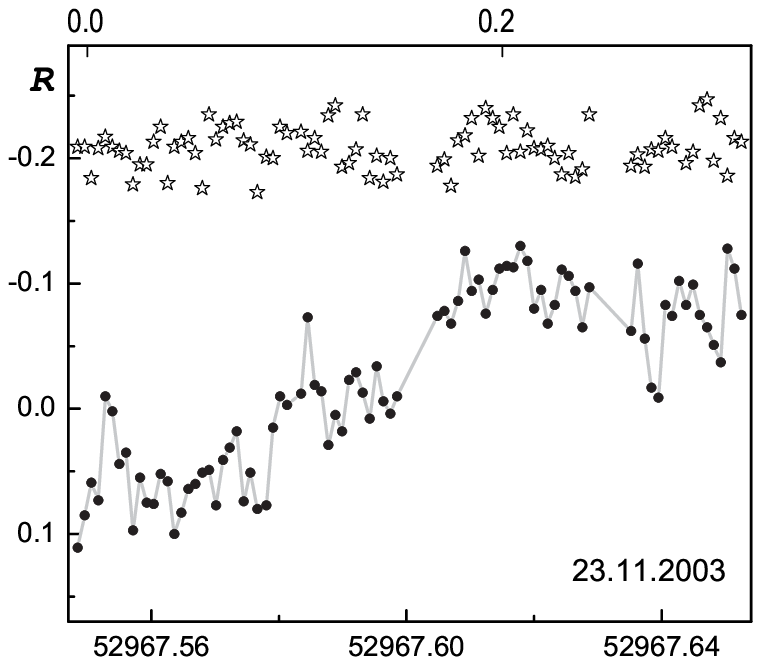} \\
\end{minipage}
\begin{minipage}[h]{0.49\linewidth}
\includegraphics[width=1\linewidth]{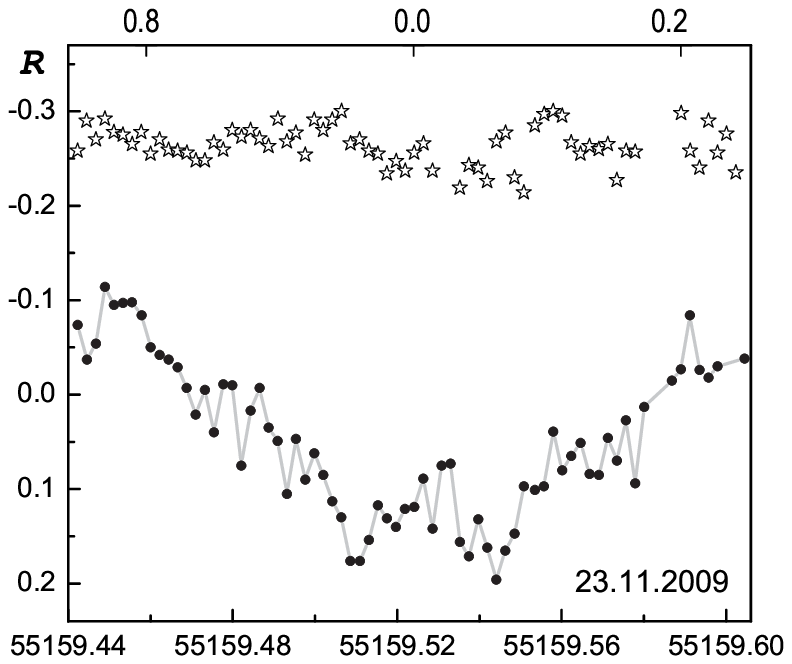} \\
\end{minipage}
\vspace{-2mm}
\vfill
\begin{minipage}[h]{0.49\linewidth}
\includegraphics[width=1\linewidth]{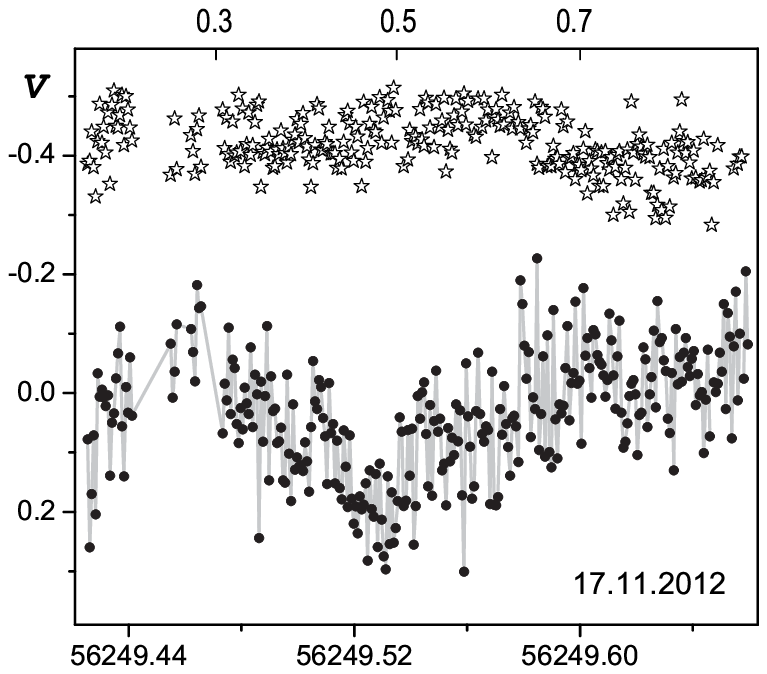} \\
\end{minipage}
\begin{minipage}[h]{0.48\linewidth}
\includegraphics[width=1\linewidth]{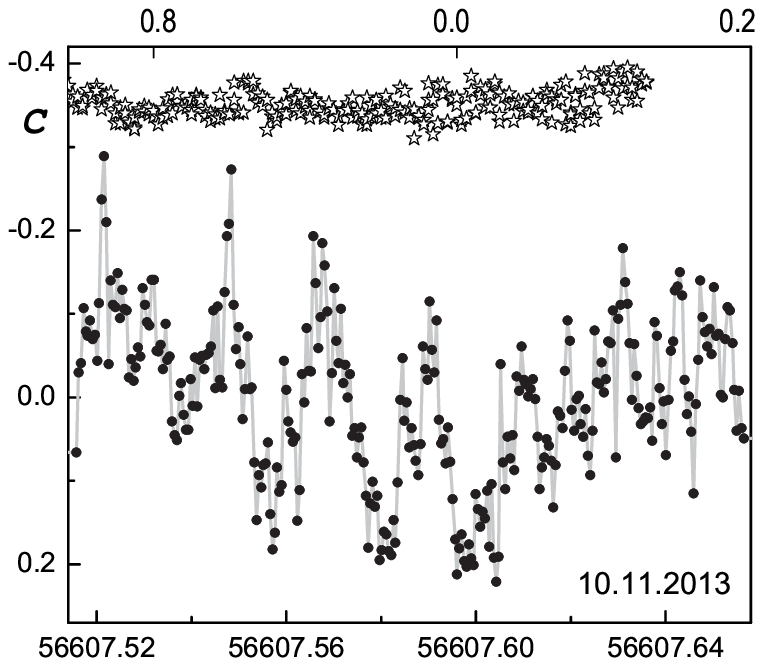} \\
\end{minipage}
\vspace{-2mm}
\vfill
\begin{minipage}[h]{0.46\linewidth}
\includegraphics[width=1\linewidth]{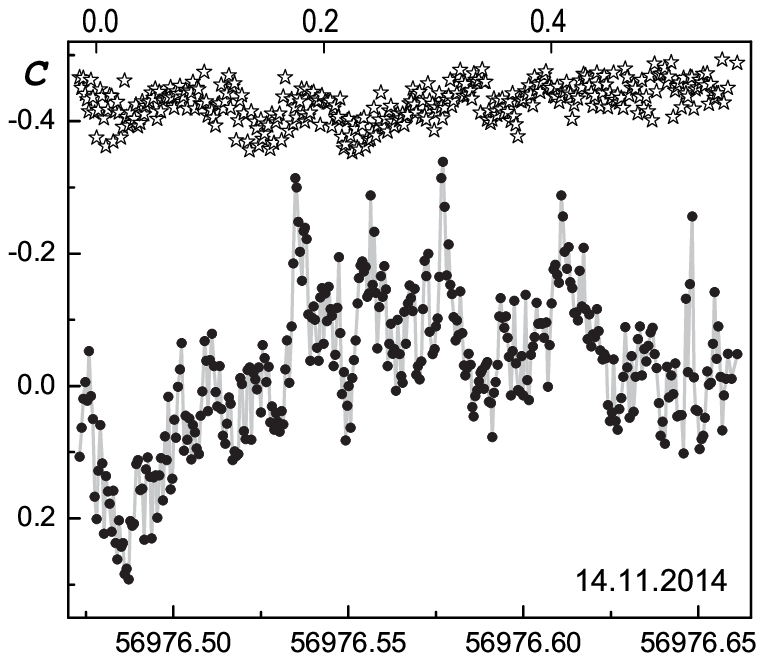} \\
\end{minipage}
\begin{minipage}[h]{0.52\linewidth}
\includegraphics[width=1\linewidth]{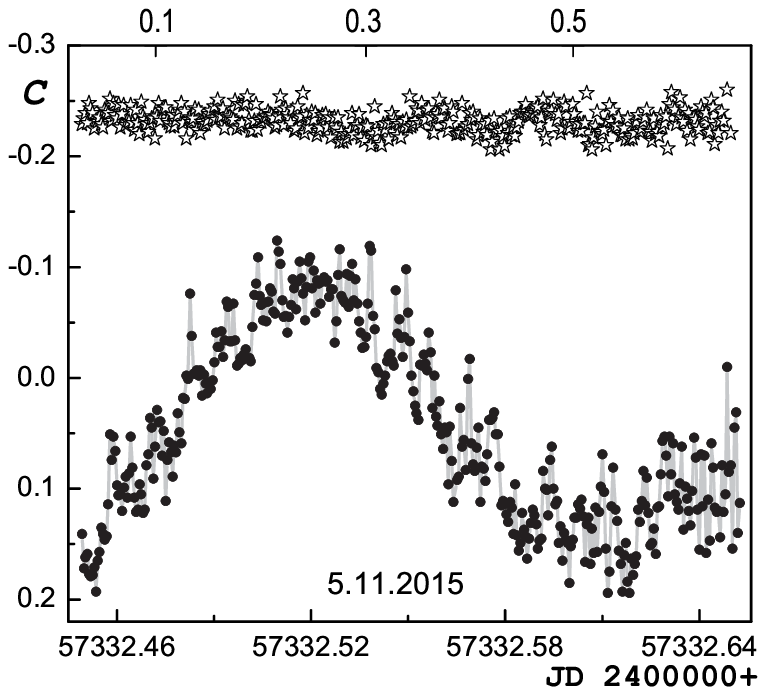}  \\
\end{minipage}
\vspace{-4mm}
\caption{The LCs for selected nights during 2003-2015. The scale of stellar magnitudes was nominated to zero. The orbital phases on the upper scale of each panel were calculated using the ephemeris (1). The night-run variability of a nearby field star of a similar brightness is shown at the top of each panel.}
\vspace{-2mm}
\end{figure}

Cantrell et al. (2010) modelled ellipsoidal light curves of V616 Mon in passive states, which occurred during the years 1988-2003, to determine the following basic parameters of the binary: {\it i} = 50$^\circ.98\pm 0^\circ.87$, $M_1$ = 6.61$\pm$0.25 $M_\odot$, $ M_2$ = 0.40$\pm$ 0.045$M_\odot$ and its distance {\it d} = 1.06$\pm$0.12 kpc. They found that all non-ellipsoidal variability originates in a variable accretion disk component. The radiation from the disk is responsible for secular changes in brightness, flickering present in non-passive states and periodic modulations in the shape of the LC. Khruzina et al. (1988) suggested the existence of a hot spot arising on the outer edge of the system. Marsh et al. (1994) and Neilsen et al. (2008) presented Doppler tomography of the accretion disk, revealing an asymmetrical disk with a prominent bright spot associated with the impact of the mass transfer stream on the edge of the disk.

Gonz\'ales Hern\'andes et al. (2014) used spectroscopic data to determine a negative orbital period derivative $dP/P$ = -0.60$\pm$0.08 ms yr$^{-1}$ from five times of inferior conjunctions of the red dwarf secondary.

Comprehensive reviews of close binaries with a black-hole component were published by Cherepashchuk (1996, 2013, 2014).

\section{The CCD observations of V616 Mon, types of variability and light elements}
In the years 2003-16 we obtained the time-resolved CCD photometry of V616 Mon in $V,R,R_C,I$ passbands and the integral light (i.e. without filters, denoted by $C$ in this paper), using the 50--125 cm telescopes at the Crimean Astronomical Station of the Lomonosov Moscow State University and Observatory Star\'a Lesn\'a of the Astronomical Institute of the Slovak Academy of Sciences. The observations in December 2015 were carried out with the 250 cm telescope, located in the Caucasian Mountain Observatory of the Sternberg Astronomical Institute of the Lomonosov Moscow State University. In total, we obtained 7660 CCD frames. The journal of observations is given in Table~1.

The LC of the mean night run observations in different passbands and integral light is shown in Fig. 1. The variability on the scale from days to months can reach 0.35 mag. The standard deviation $\sigma$ of the observations taken during every observational season, after removal the mean brightness variations, is shown in the upper part of the figure. It characterizes the activity of the system that reached its maximum in 2012.

The orbital phase LC can be characterized by a double wave, caused by an ellipsoidal shape of the red dwarf component. We used the minima times determined by our orbital LCs and the times of inferior conjunction of the red dwarf derived from spectroscopic data (Gonz\'ales Hern\'andez et al., 2014 and references therein) shifted by 0.5 in the orbital phase, to improve the light elements of the binary. For more precise spectroscopic data we applied the weight 10 times larger than for photometric ones. A period analysis of minima times allowed us to improve the linear ephemeris of the binary to:

$$ HJD(Min I) = 2457332.4400 + 0.32301407(5)\times E.\,\,\,\,\,\,\,\,\,\,\,\,(1)$$

The primary minimum corresponds to a superior conjunction of the red dwarf.
The average 2003-2016 phase LC of V616 Mon, after removal of the mean brightness variations, is depicted in Fig. 2. The standard deviation $\sigma$ is shown in the upper part of the figure.

We clearly see  that the phase LC for the whole time interval exhibits a large scatter, significantly higher than observational errors. The scatter of the data 0.2 - 0.4 mag, caused by a flickering, can increase to 0.6 mag in orbital phases 0.3 -- 0.4 and 0.8 -- 0.9. Flickering is due to the gaseous stream - accretion disk interaction in a bright spot, where the gas stream from the L1 point hits the outer edge of the accretion disk. Bisikalo et al. (2008) and Kononov et al. (2012) performed three-dimensional gas-dynamical simulations for accretion disks and found that the spot at the disk is extended and has the shape of a "hot line", or an arc. The extended hot line is best visible in the phases 0.3 -- 0.4 and, due to the favorable inclination angle of the binary {\it i} = 50$^\circ.98\pm 0^\circ.87$ (Cantrell et al., 2010), it is partly visible also in the phases 0.8 -- 0.9. The position of the observer is approximately perpendicular to the hot line in these phases, so the flickering amplitude reaches the maximum.

The LCs of V616 Mon at six nights during the years 2003 - 2015 are illustrated in Fig. 3. The observations show a superposition of the orbitally - related wave variation and a rapid flickering variability with a characteristic time scale of about 30 min and the amplitude of 0.2 - 0.4 mag.   While the orbital wave variation dominates all LCs, the largest flickering variability was detected in the years 2011-14 with its maximum in 2012. We plotted also the night run variability of the field star, with approximately the same brightness, located close to V616 Mon. The visible dispersion of the field star brightness shows the actual measurement accuracy of V616 Mon. In 2015 the amplitude of flickering decreased, so the system entered a non-active state.

\section{Discussion and conclusions}

As we have mentioned in the Introduction, the last two outbursts of V616 Mon occurred in 1917 and 1975. If the outburst is triggered by a periodic mechanism, we can expect a new outburst during the next 10 - 20 years. In 2004-6 and 2009-14 active stages we observed flickering caused by an increase of the mass transfer from the red dwarf optical component to the accretion disk of the black hole (Shugarov et al., 2015).  We have detected three types of brightness variations: mean brightness variations with a characteristic time from days to months, orbital LC variations caused by the ellipsoidal shape of the red dwarf component  and irregular flickering variations arising in the black hole accretion disk and the bright spot, where the mass transfer stream hits the outer edge of the disk. V616 Mon entered a non-active stage in the year 2015. The fast brightness variability was also detected in other X-ray nova - black hole candidates, e.g. in V404 Cyg (Pavlenko et al., 1996; Hynes et al., 2002; Shahbaz et al., 2003) and KV UMa (Gonz\'ales Hern\'andez et al., 2014; Pavlenko et al., 2001).

Flickering is observed in all types of compact binaries consisting of a late-type star transferring matter onto an accreting white dwarf, a neutron star or a black hole. Many statistical properties of this variability
can be naturally explained with the fluctuating accretion disk model, where variations in the mass-transfer rate through the disk are modulated on the local viscous timescale and propagate towards the central compact object (Scaringi, 2014). The radiation produced by accretion is emitted through a release of gravitational potential energy. Most of the radiation of accreting black holes and neutron stars is emitted in the inner parts of the accretion disk in X-rays, while optical radiation is emitted from the outer parts of the accretion disk. Intrinsic brightness fluctuations on timescales of seconds to tens of minutes in mass-exchanging binaries were reviewed by Baptista (2015). In accreting white dwarf systems most of the radiation is emitted in the optical. The flickering originates either in the inner disk region and its boundary layer and/or in the hot spot, where the mass transfer stream hits the outer edge of its accretion disk.
The LCs of classical, symbiotic and dwarf novae with a white dwarf compact object often indicate flickering, e.g.: V2468 Cyg (Chochol et al., 2012), V959 Mon (Shugarov et al., 2014), V2275 Cyg (Ostrova et al., 2003), DK Lac (Katysheva, N.A. \& Shugarov, S.Yu.), ER UMa (Zemko et al., 2014), V426 Oph (Shugarov, 1983; Hessman, 1988; Homer et al., 2004), MV Lyr (Pavlenko \& Shugarov, 1998; 1999; Katysheva et al., 1999; Pavlenko et al., 2005; Scaringi, 2014),
SW UMa (Pavlenko et al., 2000), DO Dra (Szkody et al., 2002), V795 Her (Mironov et al., 1983; \v{S}imon et al., 2012), UU Aqr (Volkov et al., 1986; Volkov \& Volkova, 2003; Baptista, 2015; Khruzina et al., 2015), RS Oph (Zamanov et al., 2015), CH Cyg (Sokoloski, 2003; Stoyanov et al., 2015, Shugarov et al., 2015) and V407 Cyg (Kolotilov et al., 2003; Shugarov et al., 2007).

A detailed description of the orbital period methodology refinement, the average LCs for each year and detailed journal of observations will be published elsewhere.

\acknowledgements
The authors thank to our referee Dr. Elena Pavlenko for constructive comments and advices.
This work was supported by the Slovak Research and Development Agency under the contract No. APVV-15-0458, the VEGA grant No. 2/0002/13, the RFBR grants No. 14-02-00825, No. 15-02-06178 and the Russian President grant NSh-9670.2016.2. A.D. acknowledges (partial) support from M.V. Lomonosov Moscow State University Program of Development.

\end{document}